\documentclass[sigplan]{acmart}

\AtBeginDocument{%
  }

\usepackage{placeins}
\usepackage[english]{babel}
\usepackage{subcaption} 

\copyrightyear{2025}
\acmYear{2025}
\setcopyright{acmlicensed}\acmConference[UCC '25]{2025 IEEE/ACM 18th International Conference on Utility and Cloud Computing}{December 1--4, 2025}{France, France}
\acmBooktitle{2025 IEEE/ACM 18th International Conference on Utility and Cloud Computing (UCC '25), December 1--4, 2025, France, France}
\acmDOI{10.1145/3773274.3774268}
\acmISBN{979-8-4007-2285-1/2025/12}
\setcopyright{none}           
\begin{document}

\title{Knowledge Graphs-Driven Intelligence for Distributed Decision Systems}

\author{Rosario Napoli}
\orcid{0009-0006-2760-9889}
\affiliation{%
  \department{Department of Mathematical and Computer Sciences, Physical Sciences and Earth Science}
  \institution{University of Messina}
  \city{Messina (ME)}
  \state{Sicily}
  \country{Italy}
}\email{rnapoli@unime.it}

\author{Gabriele Morabito}
\orcid{0009-0006-2144-8746}
\affiliation{%
  \department{Department of Mathematical and Computer Sciences, Physical Sciences and Earth Science}
  \institution{University of Messina}
  \city{Messina (ME)}
  \state{Sicily}
  \country{Italy}
}\email{gamorabito@unime.it}

\author{Antonio Celesti}
\orcid{0000-0001-9003-6194}
\affiliation{%
  \department{Department of Mathematical and Computer Sciences, Physical Sciences and Earth Science}
  \institution{University of Messina}
  \city{Messina (ME)}
  \state{Sicily}
  \country{Italy}
}\email{acelesti@unime.it}

\author{Massimo Villari}
\affiliation{%
  \department{Department of Mathematical and Computer Sciences, Physical Sciences and Earth Science}
  \institution{University of Messina}
  \city{Messina (ME)}
  \state{Sicily}
  \country{Italy}
}
\email{mvillari@unime.it}

\author{Maria Fazio}
\orcid{0000-0003-3574-1848}
\affiliation{%
  \department{Department of Mathematical and Computer Sciences, Physical Sciences and Earth Science}
  \institution{University of Messina}
  \city{Messina (ME)}
  \state{Sicily}
  \country{Italy}
}\email{mfazio@unime.it}


\renewcommand{\shortauthors}{R. Napoli, G. Morabito, A. Celesti, M. Villari, M. Fazio}

\renewcommand{\qedsymbol}{}

\begin{abstract}
  Modern distributed decision-making systems face significant challenges arising from data heterogeneity, dynamic environments, and the need for decentralized coordination. This paper introduces the Knowledge Sharing paradigm as an innovative approach that uses the semantic richness of Knowledge Graphs (KGs) and the representational power of Graph Embeddings (GEs) to achieve decentralized intelligence. Our architecture empowers individual nodes to locally construct semantic representations of their operational context, iteratively aggregating embeddings through neighbor-based exchanges using GraphSAGE. This iterative local aggregation process results in a dynamically evolving global semantic abstraction called Knowledge Map, enabling coordinated decision-making without centralized control. To validate our approach, 
  we conduct extensive experiments under a distributed resource orchestration use case. We simulate different network topologies and node workloads, analyzing 
  the local semantic drift of individual nodes.
  Experimental results confirm that our distributed knowledge-sharing mechanism effectively maintains semantic coherence and  adaptability, making it suitable for complex and dynamic environments such as Edge Computing, IoT, and multi-agent systems.
\end{abstract}

\ccsdesc[500]{Information systems~Decision support systems}

\keywords{Knowledge Sharing, Distributed Decision Systems, Knowledge Engineering, Distributed Systems}


\maketitle

\section{Introduction}\label{sec:intro}

The way we handle and organize knowledge is rapidly changing, driven by the increasing volume, velocity, and variety of data generated in modern digital systems. In particular, this explosion of information, commonly referred to as Big Data, demands new strategies that can manage not only massive amounts of data but also the intricate connections within it \cite{b1}. 
In this context, Knowledge Graphs (KGs) are a disruptive innovation for organizing semantically rich and heterogeneous data sources, as they improve basic graph structures, allowing nodes and relationships to have 
multiple labels and properties, providing an intuitive and powerful way to describe real-world dynamics \cite{b2}. 
Key components in KGs are Graph Embeddings (GEs), i.e. feature vectors representations that compress KG nodes structural and semantic information into a low-dimensional space that can be easily processed \cite{b3}.

At the same time, distributed systems have become essential for managing and processing information across geographically or logically separated nodes, especially in scenarios that require real-time responsiveness, autonomy, and high availability \cite{b4}. These systems form the basis of modern architectures such as edge computing frameworks, sensor networks, and multi-agent systems \cite{b5}.

While their decentralized nature offers advantages in terms of scalability, fault tolerance, and parallelism, it also introduces a range of non-trivial challenges. In particular, coordination among nodes often relies on partial and delayed information, leading to decision-making processes that are based on incomplete or outdated system states \cite{b6}. This is further complicated by the heterogeneity of data sources, asynchronous communication mechanisms, and dynamic changes in network topology. As a result, maintaining global consistency, ensuring knowledge alignment across nodes, and developing synchronized decision-making processes becomes a complex task. Such challenges are often solved with approaches based on centralized control or predefined consensus schemes that tend to limit scalability, making them unsuitable for highly dynamic environments \cite{b19}.

To address these limitations, this work proposes a Knowledge Graph-Driven Intelligence for Distributed Decision Systems that unifies the semantic depth of KGs with the representational power of GEs. In our architecture, we introduce the concept of \textit{Knowledge Sharing}, a new paradigm where each node semantically interprets its own local context and iteratively integrates information from neighboring nodes through GEs, forming a collective, heterogeneous, and evolving understanding of the system without any centralized coordination.

In particular, each node independently processes its raw data to construct a semantically informed representation of its local environment, encapsulated in a GE feature vector. This \textit{local knowledge} is then shared and aggregated with information from neighboring nodes through rounds of embedding propagation. Over time, this continuous interaction generates a collectively structured view of the system, leading to the construction of a unified, low-dimensional \textit{Knowledge Map}. This map acts as a semantic abstraction layer that reflects both the structural and behavioral characteristics of the system as a whole, allowing each node to reason and act not only based on its own observations, but also in alignment with all system goals. The proposed work is organized as a 4-layer architecture, specifically: the Physical Layer, responsible for sensing and data acquisition; the Storage Layer, for distributed and fault-tolerant data persistence; the Knowledge Layer, where semantic enrichment and global reasoning occur based on locally collected information; and finally, the Decision Layer, which generates autonomous and context-aware actions.

In this paper, we will detail the complete system architecture and explore the methodologies and mechanisms for knowledge generation and sharing specifically within the Knowledge Layer. To evaluate the effectiveness of the system, we conduct a set of experiments simulating real-world distributed scenarios with variable workloads and graph structures. The results demonstrate that the proposed approach achieves robust knowledge synchronization across nodes. 
These findings validate the practical viability of our architecture and open new perspectives for decentralized cognitive systems in heterogeneous and knowledge-intensive environments.

The remainder of this paper is organized as follows: Section \ref{sec:definitions} provides formal definitions of the key concepts used throughout the paper. Section \ref{sec:sota} reviews the relevant literature, highlighting the current state of the art in distributed decision-making and graph-based intelligence. 
Section \ref{sec:arch} presents the proposed system architecture, detailing its four functional layers and their respective responsibilities. Section \ref{sec:knowledge} focuses on the Knowledge Sharing mechanism, explaining how local and global embeddings are propagated and aggregated across the network.
Section \ref{sec:usecase} describes a use case centered on the distributed orchestration of services across a dynamic computing infrastructure, while Section \ref{sec:test} outlines the experimental setup and evaluates the performance of the Knowledge Sharing mechanism based on the presented use case under different conditions. Finally, Sections \ref{sec:results_discussion} and \ref{sec:conclusions} conclude the article and suggest directions for future research.

\section{Definitions}\label{sec:definitions}

This section establishes the formal definitions of the key concepts adopted in this work.
\begin{definition}[Knowledge Graph]
A Knowledge Graph is formally defined as a tuple \( KG = (V, E, R, \lambda_V) \) \cite{kgdef}, where:
\begin{itemize}
    \item \( V \) represents the collection of entities (nodes);
    \item \( R \) denotes the set of relationships types;
    \item \( E \subseteq V \times R \times V \) is the set of directed edges;
    \item \( \lambda_V : V \rightarrow 2^T \) is a classification function that assigns one or more type labels from the set \( T \) to each node.
\end{itemize}
\end{definition}

\begin{definition}[KG atomic unit]
The atomic unit of a KG is a triple \cite{b7}, composed of:
\begin{itemize}
    \item  a typed source node \( s \);
    \item a typed target node \( t \);
    \item a directed typed edge \( r \) connecting \( s \) to \( t \).
\end{itemize}
\end{definition}
\begin{definition}[Graph Embedding]
Let \( KG = (V, E, R, \lambda_V) \) be a Knowledge Graph. A Graph Embedding (GE) is a function
\[
f: V \rightarrow \mathbb{R}^k,
\]
where \( k \ll |V| \), which associate each node \( v \in V \) to a vector \( f(v) \in \mathbb{R}^k \) \cite{b8}.
\end{definition}

\begin{definition}[GraphSAGE on Knowledge Graphs]
Given a Knowledge Graph \( KG = (V, E, R, \lambda_V) \), GraphSAGE (Graph Sample and Aggregate) is an inductive framework designed to learn embeddings by collecting and combining information from the immediate vicinity of each node \cite{b27}.
At each iteration \( l \), the representation of a node \( v \in V \), denoted as \( z_v^{(l)} \in \mathbb{R}^k \), is updated as follows:
\begin{multline}
z_v^{(l)} = \sigma\Big( U^{(l)} \cdot
\text{AGG}^{(l)}\big( \{ z_u^{(l-1)} \mid (u, r, v) \in R \} \\
\cup \{ z_v^{(l-1)} \} \big) \Big)
\end{multline}
where:
\begin{itemize}
    \item \( z_v^{(0)} \) represents the initial feature vector for node \( v \);
    \item \( \text{AGG}^{(l)} \) is a permutation-invariant aggregation function;
    \item \( U^{(l)} \) is a trainable weight matrix for iteration \( l \);
    \item \( \sigma \) is a non-linear activation function.
\end{itemize}

GraphSAGE facilitates inductive learning, allowing for generalization to previously unobserved nodes through learned aggregation functions applied over sampled neighborhoods \cite{b9}.
\end{definition}

\section{Related Work}\label{sec:sota}
KGs have become increasingly important in addressing the complexity of data integration and semantic representation in modern computing paradigms \cite{b20}. In Edge Computing and Internet of Things (IoT) environments, where data is inherently distributed and heterogeneous, KGs offer a promising abstraction layer capable of semantically organizing different data sources and enabling localized reasoning \cite{b10}. Their ability to unify data from spatially sparse sensors and devices into a coherent  structure is well established, with solutions, such as SuccinctEdge, that have demonstrated that even resource-constrained nodes can benefit from compact, in-memory KG representations designed for fast and decompression-free querying of 
data \cite{b11}.

Semantic access control has also gained attention for its ability to enforce context-aware policies, where attribute-based rules provide flexible enforcement mechanisms in domains like smart homes and sensitive environments \cite{b16}.
In this context, Security-related reasoning systems based on KGs and ontologies, like IoTSTO, have shown how semantic modeling can be applied to decision making processes related to security analysis \cite{b14}. Nevertheless, scalability and the need for fast updates in dynamic environments remain major challenges. 
Similar limitations affect self-healing and malfunction detection architectures that adopt KGs for monitoring distributed systems. Several frameworks have tried to address semantic challenges through modular KG architectures that integrate heterogeneous sources such as BIM models, sensor metadata, and operational processes \cite{b13}, providing flexible models for evolving data landscapes. Hovewer, these methods rely on big data analytics and probabilistic models for autonomous fault recovery, struggling with synchronization and consistency across heterogeneous nodes to make efficient decisions \cite{b15}.

From a learning perspective, Graph Embedding (GE) techniques have been widely employed to derive compact and expressive representations of graph-structured data \cite{b24}. However, these embeddings are usually computed either in centralized settings or based on static graph snapshots that only used as input for ML models. Their application in evolving and distributed KGs, especially under edge constraints, is still limited. Most existing models are not designed to reflect frequent structural changes or the real-time evolution of KG topologies \cite{b2}, and embedding updates tend to struggle to keep up data streaming. Moreover, Ontology-driven models enrich embedding context-sensitive policies, particularly for  decisions-making processes in domains such as security and access control \cite{b23}. However, the complexity of ontologies and the latency introduced by rich axiomatic reasoning (i.e., reasoning from self-evident truths) limit their viability in real-time deployments at the Edge, with efforts that are still at a formative stage. In the end, recent efforts are moving toward lightweight, incremental embedding strategies \cite{b4}, but their integration into  distributed knowledge reasoning architectures remains unexplored.

As a result, the convergence of local semantic awareness, global embedding propagation, and decentralized decision-making is still in its beginning.  In current literature, many existing approaches focus on enhancing isolated capabilities, such as semantic reasoning, data integration, or context awareness, 
but rarely consider how these elements might interact and evolve across nodes in a system where no central coordination exists. Even in systems with modular KGs, semantic enrichment and inference are typically performed in a localized way, without any notion of collective knowledge emergence through network-wide interactions. In particular, existing systems often exchange raw or partially structured data, leading to fragmented reasoning and inconsistent decisions \cite{b30}.
Standardization efforts for cross-domain ontologies and semantic interoperability are ongoing \cite{b18}, but the lack of cohesive frameworks that unify local and global knowledge perspectives in distributed environments is evident, revealing a significant gap in the state of the art.

While centralized or modular KG architectures have been applied in the Edge and IoT domains \cite{b21}, the idea of enabling each node with a local shard of a distributed KG and enabling decentralized cooperation based on semantic embeddings and shared knowledge propagation is still missing from the current landscape. The challenges of heterogeneity in device capabilities, inconsistent data formats, and asynchronous communication continue to limit the emergence of a unified, distributed semantic intelligence.

In particular, the integration of KGs into truly distributed decision-making systems remains a novel and only partially explored area. 

This fragmented scenario suggests that the time for a  paradigm shift is coming: one in which each node contributes to and benefits from a shared semantic ecosystem, dynamically shaped by localized embeddings and collective reasoning without relying on centralized coordination. The principles of knowledge locality, sharing, and embedding refinement must be rethought from a system-wide perspective to meet the growing demands of adaptability, autonomy, and scalability in next-generation distributed systems.

\section{System Architecture}\label{sec:arch}
To address the lack of a real semantic knowledge management in distributed systems, the absence of a decentralized information sharing, and difficulties in maintaining updated and synchronized KGs representations in dynamic distributed environments, a paradigm shift in the design of distributed decision-making systems is required.

{Our approach specifically aims to bridge these gaps by combining the semantic strength of KGs with innovative and distributed GEs techniques. In particular, we propose a 4-layer architecture (Fig. \ref{fig:arch}) that enables each system node to locally construct and continuously refine a semantic representation of its own data, enriching it iteratively through a decentralized knowledge-sharing process. Specifically, each node actively contributes to the generation of a shared global view, giving adaptability to real-time changes and overcoming the limitations of traditional methods that rely on heavy ontologies or static and centralized embedding techniques.

In the following, we will describe our system architecture in detail, emphasizing how each of the proposed functional layers specifically addresses these challenges, resulting in a game-changer framework for intelligent knowledge management in distributed systems.

\begin{figure}[h]
    \centering
    \includegraphics[width=0.7\linewidth]{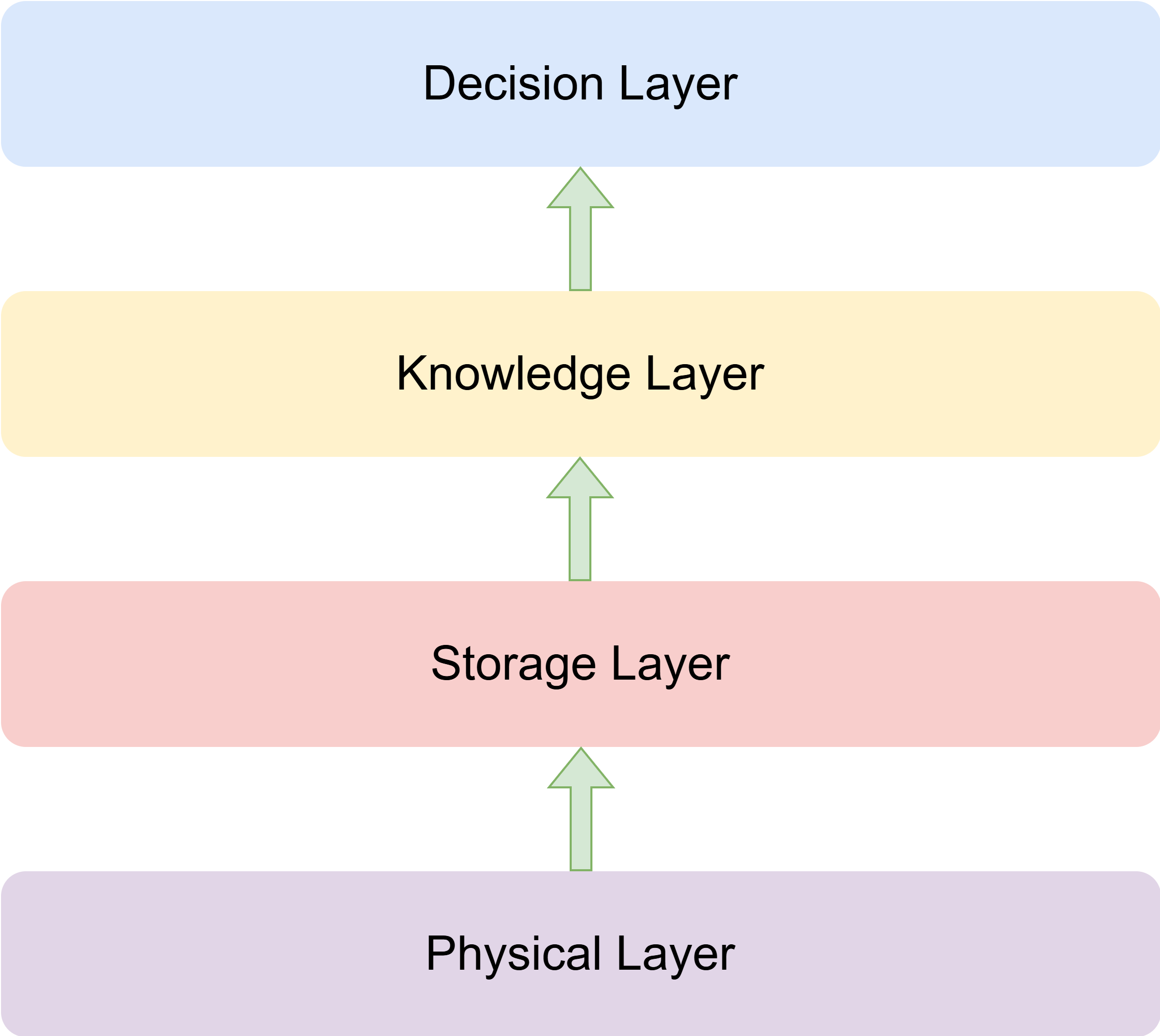}
    \caption{High-level overview of System Architecture}
    \label{fig:arch}
\end{figure}

\subsection*{Physical Layer}
The Physical Layer represents the physical nodes that form the distributed system. Each node within this layer is responsible for the continuous collection of raw local input data generated within its specific context. This data may include, but is not limited to, performance metrics, sensor states, system events, or any other pertinent contextual information essential for the decision-making process. 
The inherent distributed nature of this layer requires parallel and asynchronous data acquisition, which is fundamental to establishing a comprehensive and real-time systemic awareness. Raw data collected by individual nodes in the Physical Layer is sent to the Storage Layer. 

\subsection*{Storage Layer}
The Storage Layer is responsible for the persistence and the management of the data received from the Physical Layer. Given the substantial volume, high velocity, and inherent heterogeneity characteristic of data generated by distributed systems, coupled with the critical requirement for continuous operation, this layer should be implemented as a distributed storage. While a centralized storage solution might offer simplicity for smaller-scale or non-critical deployments, its adoption in a distributed decision system introduces a single point of failure (SPOF). The failure of a centralized storage component would render the entire decision-making process inoperable, severely compromising system resilience and availability. Consequently, a distributed paradigm for data management is mandated. Such an architecture typically leverages advanced features found in NoSQL database management systems, including data mirroring for fault tolerance and sharding for horizontal scalability across multiple nodes \cite{b25}. Graph-oriented databases, in particular, are well suited for modeling complex inter-data relationships and facilitating advanced pattern discovery and semantic querying, making them a strong candidate for this layer's core technology \cite{b26}.

\subsection*{Knowledge Layer}
The Knowledge Layer is responsible for transforming raw and structured data from the Storage Layer into semantically meaningful representations, expressed as Knowledge Graph Embeddings (KGEs). This transformation enables each node to interpret its local context while progressively integrating information from its neighbors. Through iterative aggregation, local knowledge becomes the basis of a global system view, resulting in embeddings that are both semantically rich and contextually coherent. The global knowledge is then represented in the knowledge map, which is the foundation for the decentralized decision-making processes. The mechanisms and methodologies for knowledge generation within this layer will be detailed in Section \ref{sec:knowledge}. 
The precise formalization of the knowledge map, which synthesizes the distributed semantic representations produced within the Knowledge Layer, is provided below.
\begin{definition}[Knowledge Map]
Given a distributed system where each node is associated with a local embedding vector $\mathbf{h}_i^{(k)}$, iteratively refined via neighbor-based aggregation (as defined for GraphSAGE in Section 2 of the paper), the \textit{Knowledge Map} is the emergent global representation:
\[
\mathcal{K}_M = \left\{ \mathbf{h}_i^{*} \mid \mathbf{h}_i^{*} = \lim_{k \to \infty} \mathbf{h}_i^{(k)},\; i \in V \right\}
\]
where $V$ is the set of nodes in the underlying KG (see Def. 2.1), and $\mathbf{h}_i^{(k)}$ are embeddings computed according to the embedding definition and aggregation equation provided in the paper.

The Knowledge Map $\mathcal{K}_M$ thus constitutes a low-dimensional manifold in which each vector encodes both the semantic and structural context of the corresponding node, as learned and shared through the distributed aggregation process. Within this space, nodes with similar semantics and roles will be geometrically proximate:
\[
\| \mathbf{h}_i^{*} - \mathbf{h}_j^{*} \|_2 \ll 1 \iff v_i, v_j \text{ are semantically close in the KG}
\]

The Knowledge Map acts as a dynamic semantic abstraction layer, continuously updated in response to local/environmental changes, and supports decentralized reasoning and decision-making across the system.
\end{definition}

\subsection*{Decision Layer}
The Decision Layer is responsible for making informed and autonomous decisions based on the refined knowledge provided by the Knowledge Layer. This layer processes embeddings and aggregated knowledge representations as its primary input. Decision generation can be achieved through the application of advanced artificial intelligence algorithms, machine learning models, or sophisticated expert systems, which collectively derive optimal actions or system configurations. The high-level decisions or refined policies disseminated by this layer enable individual nodes to make autonomous, localized choices that align with the global system objectives. This completely distributed architectural pattern for decision-making inherently enhances system responsiveness by eliminating the need for communication with a central authority and effectively distributes the cognitive load across the network, thereby avoiding bottlenecks and improving overall system resilience.

\subsection{In-depth Full Architecture Workflow}
After providing a high-level overview of the proposed architecture and highlighting its key conceptual components, it is now essential to delve into a more detailed examination of the workflow that orchestrates data, knowledge, and decision-making across the system. In the following, we present a comprehensive breakdown of the architecture's operational flow (Fig.~\ref{fig:complete_workflow}).}
\begin{figure*}[ht!]
    \centering
    \includegraphics[width=1\linewidth]{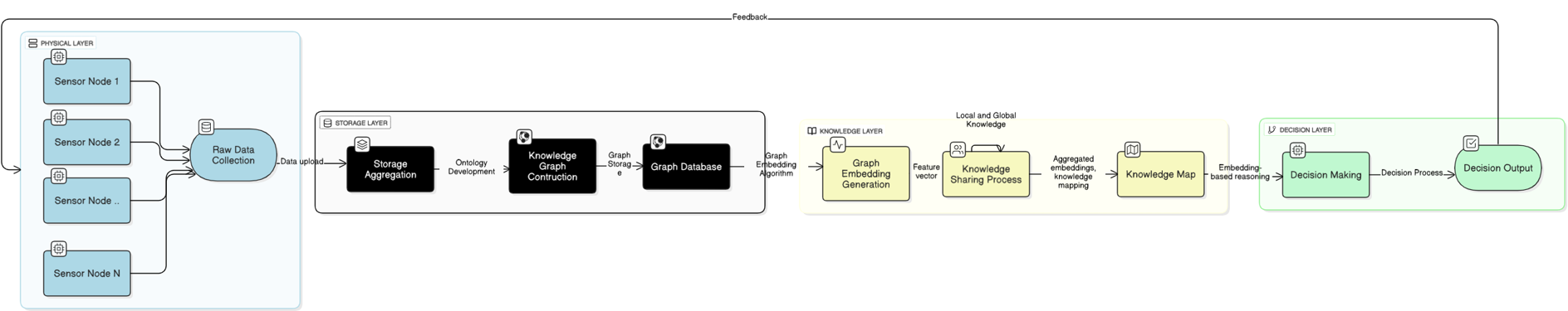}
    \caption{Full Architecture Workflow}
    \label{fig:complete_workflow}
\end{figure*}

The workflow begins at the \textbf{Physical Layer}, where multiple heterogeneous Sensor Nodes (1 to N) continuously perform Raw Data Collection, capturing  environmental and system-specific metrics. The collected data is uploaded to the \textbf{Storage Layer}, where Storage Aggregation ensures distributed and fault-tolerant persistence, preparing the data for semantic modeling.

Next, an Ontology Development phase defines the semantic schema, enabling the Knowledge Graph Construction process to map raw data into a structured graph of entities and relationships. This KG is stored in a Graph Database, which provides the necessary infrastructure for efficient Graph Storage and retrieval operations.

In the \textbf{Knowledge Layer}, the structured graph data is transformed into low-dimensional Feature Vectors through a Graph Embedding Algorithm. These embeddings capture both node attributes and contextual relationships, serving as the foundation for the Knowledge Sharing Process. Through iterative, neighbor-based exchanges, nodes propagate and aggregate embeddings, leading to the emergence of a global Knowledge Map, a dynamic semantic abstraction that reflects the system’s evolving state.

The \textbf{Decision Layer} leverages this Knowledge Map to perform Embedding-based Reasoning. The Decision Making module interprets the embeddings to derive context-aware decisions, which are materialized as actionable outputs in the Decision Output phase. These decisions are fed back into the Physical Layer, closing the loop and ensuring continuous adaptation of sensor node behavior in response to environmental changes and system dynamics.

This end-to-end workflow enables a fully distributed knowledge refinement cycle, where local data is semantically enriched, shared across nodes, and transformed into informed decisions without relying on centralized coordination.

\section{Knowledge Sharing}\label{sec:knowledge}

Building upon the functionalities of the Knowledge Layer, Knowledge Sharing represents the core mechanism that enables semantic enrichment and distributed decision support. It is a decentralized and iterative process where each node updates its internal knowledge by aggregating information from its neighbors (Fig. \ref{fig:knowledge_sharing}). This process starts from a \textit{local sharing} phase, where nodes exchange information only within their immediate neighborhood. Through successive iterations, these localized interactions propagate throughout the network, giving rise to a \textit{global sharing} effect where nodes gradually acquire a system-wide semantic awareness.

The foundation of this mechanism is the continuous construction and refinement of the KG representation of the distributed system, where entities and relationships are semantically encoded. From these evolving KGs, nodes derive KGEs, projecting structural and semantic information into low-dimensional vectors. These embeddings are exchanged and used for orchestrating decisions and coordinating actions without the need for centralized control, ensuring that each node contributes to a shared and coherent understanding of the system's state.

\subsection{Knowledge Embedding and Feature Aggregation}

{Knowledge Sharing lies in the KGs ability to transform complex graph-based information into structured, comparable representations. This is accomplished through the generation of KGEs, which convert the heterogeneous topology of a KG, with nodes and relationships properties into a low-dimensional vector space where both semantic meaning and structural patterns are preserved

In our architecture, KGEs are not static; they are continuously updated as new data becomes available and as nodes engage in feature exchange with their neighbors. Each node thus refines its own representation by integrating knowledge derived from its immediate environment and, over time, from the broader system. This process enables the construction of a \textit{unified knowledge map}, where nodes with similar roles, behaviors, or semantics are geometrically proximate.

\begin{figure}[ht]
    \centering
    \includegraphics[width=0.95\linewidth]{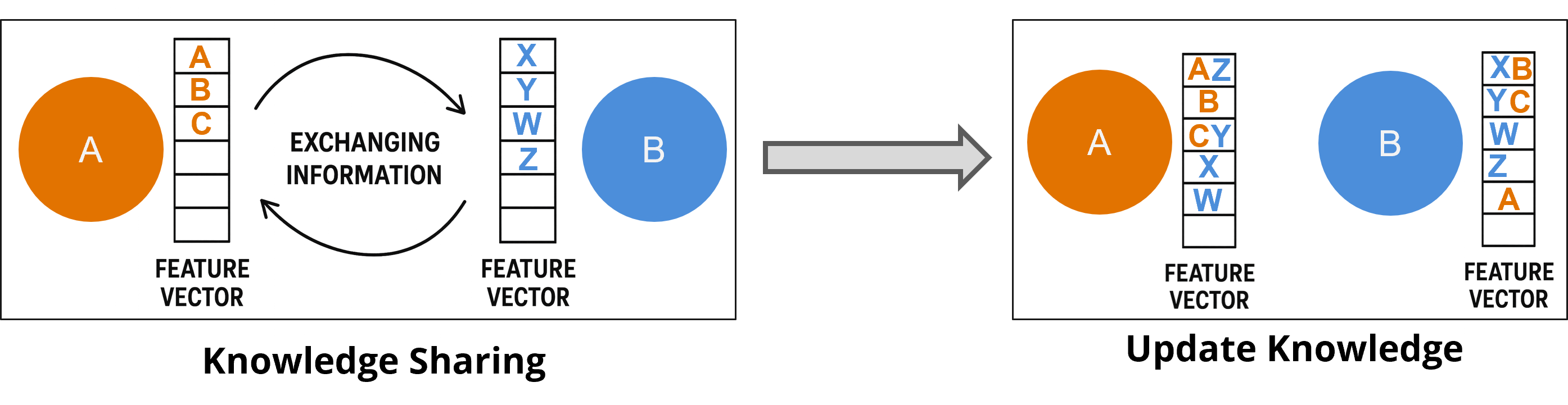}
    \caption{Distributed Knowledge Sharing between nodes A and B. Each node aggregates its feature vector with its neighbor’s information.}
    \label{fig:knowledge_sharing}
\end{figure}

The embedding process relies on \textit{feature aggregation}, where each node combines its current feature vector with those of its neighbors. 
The embeddings are therefore compact (\textit{low-dimensional}), meaningful (\textit{semantically expressive}), and actionable (\textit{usable by downstream models}).

These vector representations serve a dual purpose in our system:
\begin{itemize}
    \item They enable informed decisions within the Decision Layer, allowing nodes to reason about their role in relation to the overall system state.
    \item They act as interoperable interfaces between distributed components, supporting tasks such as anomaly detection, resource allocation, and task scheduling.
\end{itemize}

In the following Subsections, we describe how these embeddings are progressively constructed and refined through both \textit{local} and \textit{global} knowledge exchanges, supported by the GraphSAGE framework introduced in Section~\ref{sec:definitions}.

\subsection{Local Sharing}

At the foundation of the Knowledge Sharing process lies \textbf{Local Sharing}, where each node interacts only with its immediate neighbors. This exchange of information is both \textit{adaptive} and \textit{context-aware}. Each node updates its feature vector by incorporating information received from nearby nodes, while preserving its own semantic and state. This decentralized, neighbor-based communication forms the basis of scalability and responsiveness in the proposed architecture. As each node continues to update its internal knowledge with inputs from its local neighborhood, the system exhibits emergent behavior that scales efficiently with the number of nodes as will be demonstrated in Sec. \ref{sec:test}.

\subsection{Global Sharing}
As multiple local sharing domains (Fig.~\ref{fig:local_sharing}), defined by node neighborhoods, exchange information iteratively, \textbf{Global Sharing} emerges as the cumulative result of repeated Local Sharing steps.
\begin{figure}[ht]
    \centering
    \includegraphics[width=0.55\linewidth]{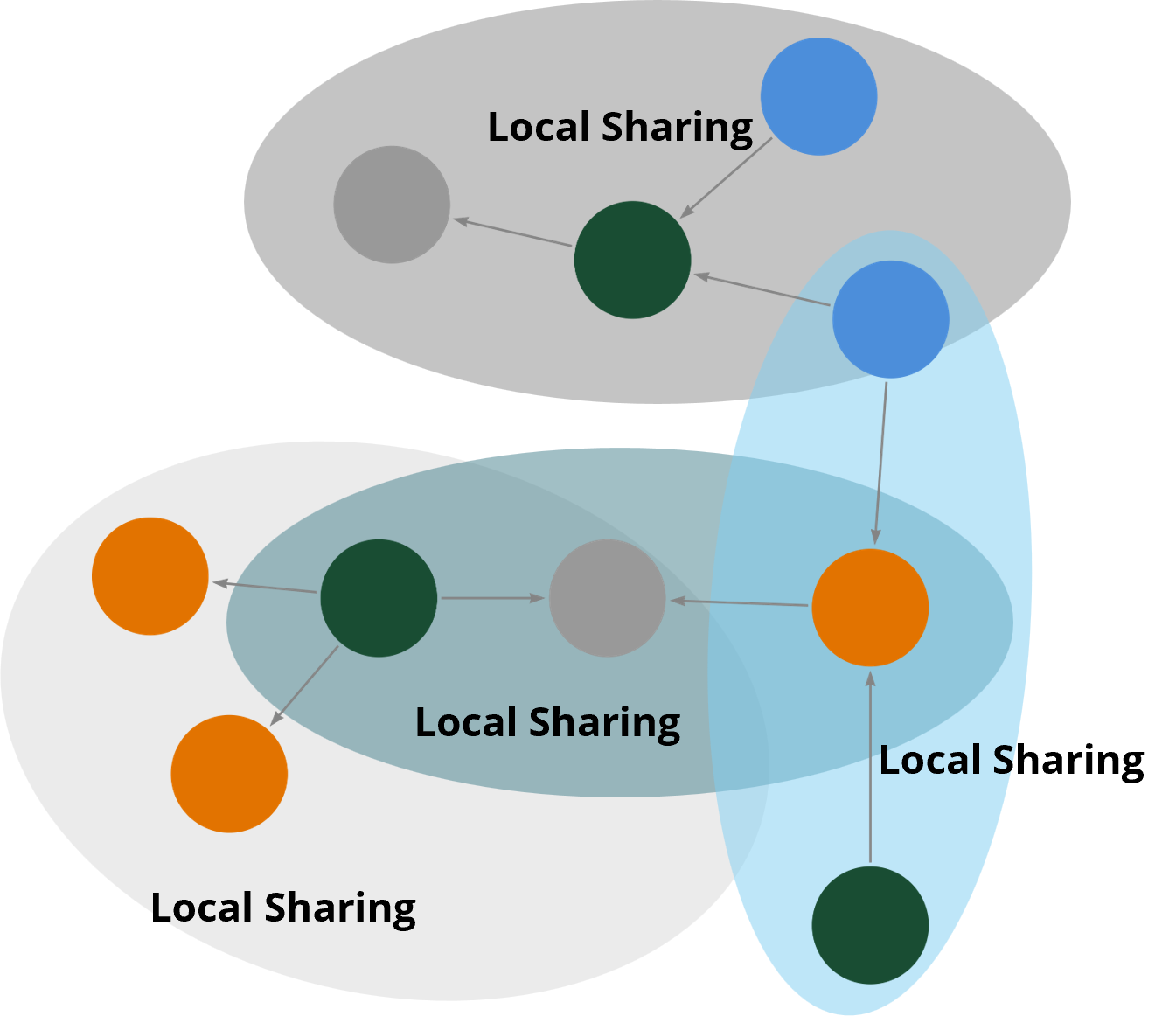}
    \caption{Local knowledge exchange between overlapping node neighborhoods enables information propagation at global level.}
    \label{fig:local_sharing}
\end{figure}

Although each node initially exchanges information only with its direct neighbors, the iterative application of local updates leads to the \textit{gradual diffusion of knowledge} across the entire graph. Through this process, nodes indirectly acquire information about distant regions of the graph, enabling reasoning that spans multiple hops. This effect yields a globally coherent view of the system that enhances decision-making processes without requiring explicit global synchronization. Global Sharing is thus the key to maintaining both autonomy and coordination in a fully distributed architecture. Nodes operate based on their local observations, yet collectively converge toward a shared understanding of the system.

\subsection{Knowledge Map Construction}

The continuous exchange of information among nodes, through both local and global sharing, gives rise to a dynamic and semantically structured representation known as the \textbf{Knowledge Map}. Unlike a static repository, this map is continuously evolving, reflecting the most recent state of collective knowledge across the entire distributed system.

Each node participates in this construction process by sharing its local perspective, which is encapsulated in its embedding vector. These vectors are not isolated descriptors but contain both intrinsic node information and contextual data gathered through neighborhood interactions. As knowledge propagates and embeddings are refined through successive aggregation steps, a consistent geometric structure emerges in the embedding space. Within this space, nodes that play similar roles, operate under similar conditions, or share semantic characteristics tend to be positioned close to one another. The result is a layout that captures not only structural relationships but also functional behavior and environmental context.

The Knowledge Map thus becomes a semantic abstraction layer for the system. It allows each node to act based on its own localized information while still adhering to a coherent global view. This mechanism enhances the system’s capacity for self-organization, coordination, and optimization. In distributed environments, where central coordination is impractical or undesirable, the Knowledge Map enables indirect but reliable global reasoning.

\subsection{The Knowledge Cycle}

The complete Knowledge Sharing mechanism is part of a continuous cycle (Fig.~\ref{fig:cicle}) in the proposed architecture, which ensures that the system remains adaptive and self-aware as it evolves over time:

\begin{enumerate}
    \item \textbf{KG Construction:} Entities and relationships are modeled as a semantic graph.
    \item \textbf{Knowledge Sharing:} Nodes exchange and aggregate information via GraphSAGE.
    \item \textbf{Embedding Propagation:} Feature vectors are refined through iterative updates.
    \item \textbf{System Adaptation:} Embeddings inform decisions at the node and system levels.
    \item \textbf{Reiteration:} New data triggers further knowledge refinement.
\end{enumerate}
\begin{figure}[ht!]
    \centering
    \includegraphics[width=0.75\linewidth]{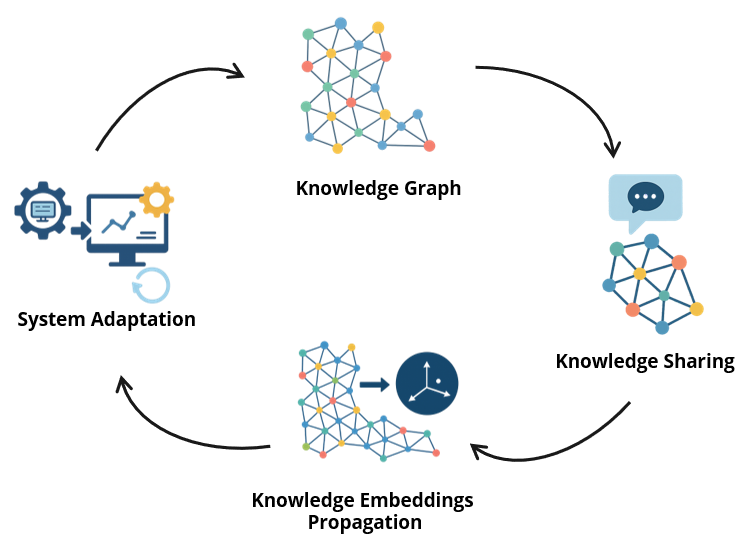}
    \caption{Knowledge Cycle.}
    \label{fig:cicle}
\end{figure}

\section{Use Case}\label{sec:usecase}
To evaluate the practical applicability of the proposed architecture, we consider a representative use case centered on the distributed orchestration of services and processes across a dynamic and heterogeneous computing infrastructure. This scenario captures the complexity of modern Edge, IoT, and multi-agent systems, where autonomous nodes must continuously collect, share, and act upon context-aware knowledge to maintain system-level objectives such as load balancing, fault tolerance, and efficient resource utilization.

The system consists of a network of distributed nodes (Physical Layer), each representing a physical or virtual entity capable of monitoring its local operational environment.
Nodes collect raw metrics such as CPU usage, memory availability, and workload status.

This data is locally processed and stored into a distributed NoSQL graph-oriented database (Storage Layer), which is deployed directly on the nodes themselves. Each node hosts a local shard of the distributed KG, embedding its own view of the system. In this model, each node in the distributed infrastructure corresponds to a KG node, enriched with properties derived from its local metrics. Edges in the KG represent communication capabilities or logical relationships between nodes, defining the actual network topology.

At the Knowledge Layer, each node independently generates a semantically enriched representation of its knowledge in the form of Graph Embeddings. These embeddings are shared with neighboring nodes through the Knowledge Sharing mechanism.  As each node refines its embeddings through iterative interactions with its neighbors, the system as a whole converges toward a globally coherent understanding of its state, enabling the progressive construction of a Knowledge Map.

In the Decision Layer, each node leverages its refined embedding to perform inference and make orchestration decisions autonomously. The decision-making process is therefore both decentralized and informed by a rich semantic understanding of the broader system context, derived through the collective knowledge propagation process.

This approach enables a fully distributed orchestration model, where nodes dynamically adapt their behavior in response to evolving conditions, without requiring any centralized coordination. Each node acts based on its local knowledge while remaining aligned with the system's global objectives through semantic consensus emerging from embedding synchronization.

\section{Experiments}\label{sec:test}
This section aims to validate the foundational concept of \textit{Knowledge Sharing}, which supports our distributed decision-making framework.
To validate our model we 
address its empirical effectiveness.

To this end,  we conduct a series of \textit{experimental validations} designed to empirically observe and quantify the semantic behavior of node embeddings under different scenarios. The experimental validations presented in this section are grounded on the use case introduced in Section \ref{sec:usecase}, which models a heterogeneous and dynamic distributed infrastructure for autonomous service orchestration. This scenario serves as the operational context for evaluating the effectiveness of the proposed Knowledge Sharing paradigm.

The experimental validation is further articulated through a complementary investigation computed by using different topologies, dimensions and workloads conditions. It focuses on \textit{Semantic Drift of a Target Node as Local Knowledge Validation}, where we track the evolution of a node's embedding as its workload varies, assessing how local semantic perceptions are preserved and transformed. 

Through this approach, we aim to establish a robust validation of the Knowledge Sharing paradigm, demonstrating its effectiveness in enabling distributed nodes to collaboratively construct a semantically rich and dynamically evolving representation of the system's state, without relying on centralized coordination.

\subsection{Experimental Setup}

To simulate a distributed scenario, we constructed a set of experiments on topologies with increasing size, ranging from 5 up to 20 nodes.
All experiments were conducted on top of \textbf{Neo4j}\footnote{neo4j.com}, a native graph database that supports the \textit{Property Graph Model}, enabling the representation of knowledge in the form of semantically enriched entities and relationships. Neo4j’s declarative query language, Cypher, and its \textbf{Graph Data Science (GDS)} library provide the foundational infrastructure for graph construction and computation. In particular, the GDS implementation of \textbf{GraphSAGE} is employed to generate embeddings that incorporate both the node’s own attributes and the aggregated information from its topological neighborhood.
Each node is modeled as a ComputationalNode in Neo4j, and is characterized by attributes such as CPU usage, memory availability, and total memory capacity. The nodes are interconnected according to three different topologies and are made in KG form, in particular:
\begin{itemize}
    \item Ring (Figure~\ref{fig:topology1});
        \item Fully Connected (Figure~\ref{fig:topology2}).
    \item Line (Figure~\ref{fig:topology3});

\end{itemize}

\begin{figure}[ht!]
    \centering
    \begin{subfigure}[b]{0.32\linewidth}
        \centering
        \includegraphics[width=\linewidth]{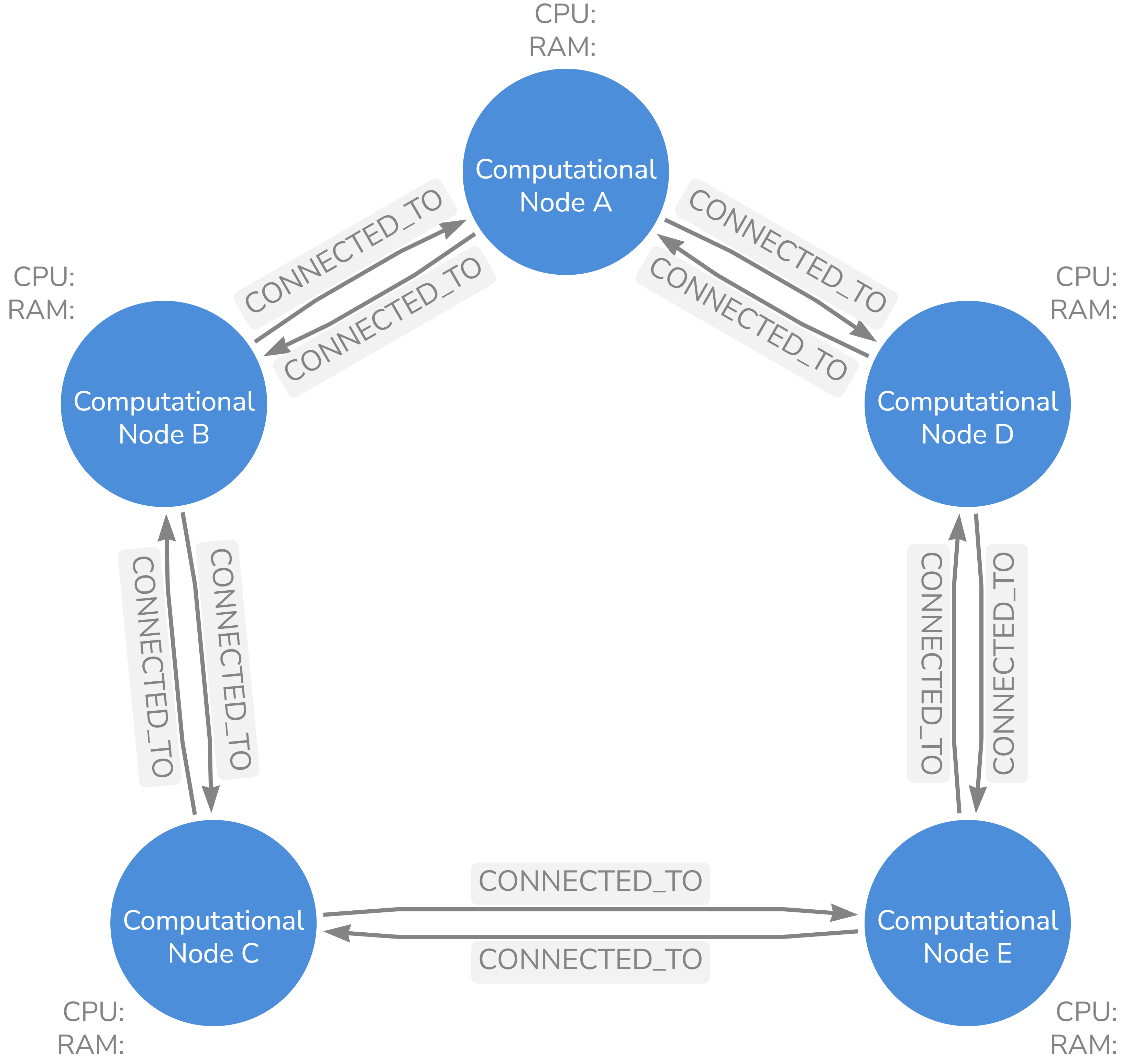}
        \caption{Ring.}
        \label{fig:topology1}
    \end{subfigure}
    \begin{subfigure}[b]{0.32\linewidth}
        \centering
        \includegraphics[width=\linewidth]{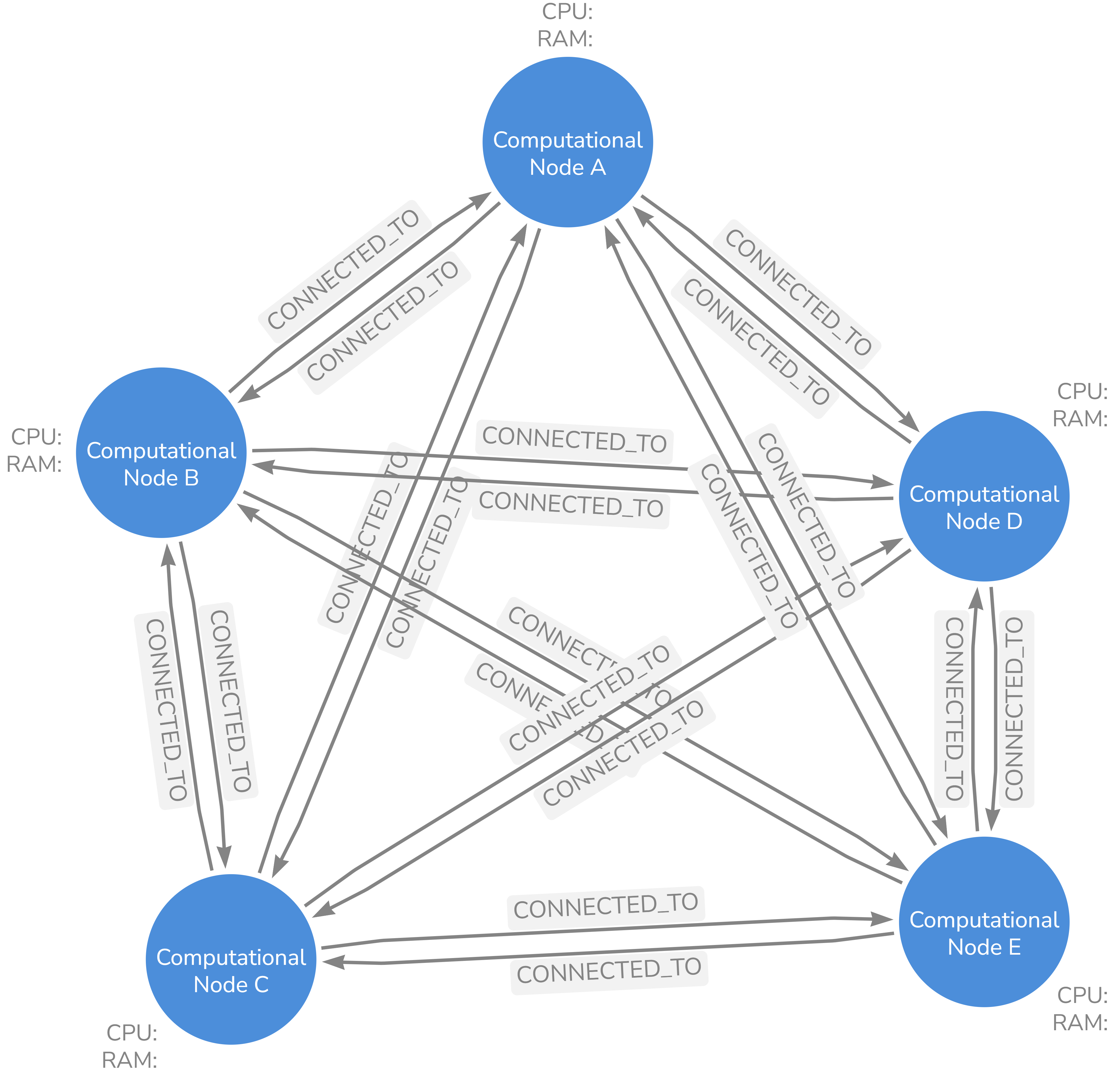}
        \caption{Fully Connected.}
        \label{fig:topology2}
    \end{subfigure}
    \begin{subfigure}[b]{0.32\linewidth}
        \centering
        \includegraphics[width=\linewidth]{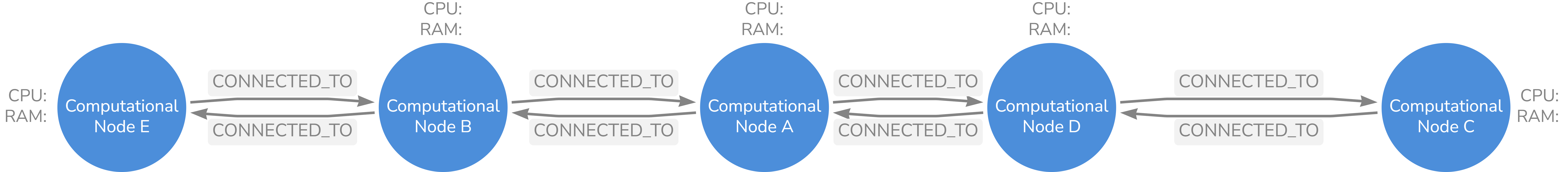}
        \caption{Line.}
        \label{fig:topology3}
    \end{subfigure}
    \caption{Overall Caption for All Topologies in KGs form.}
    \label{fig:all_topologies}
\end{figure}

In each experiment, the computational load of the nodes is progressively increased in steps of 10\% (from 10\% to 100\%), and a dedicated fluctuation engine is used to introduce slight random variations in the node RAM and CPU parameters (Fig. \ref{fig:fluctuations}).
\begin{figure}[ht!]
    \centering
    \includegraphics[width=1\linewidth]{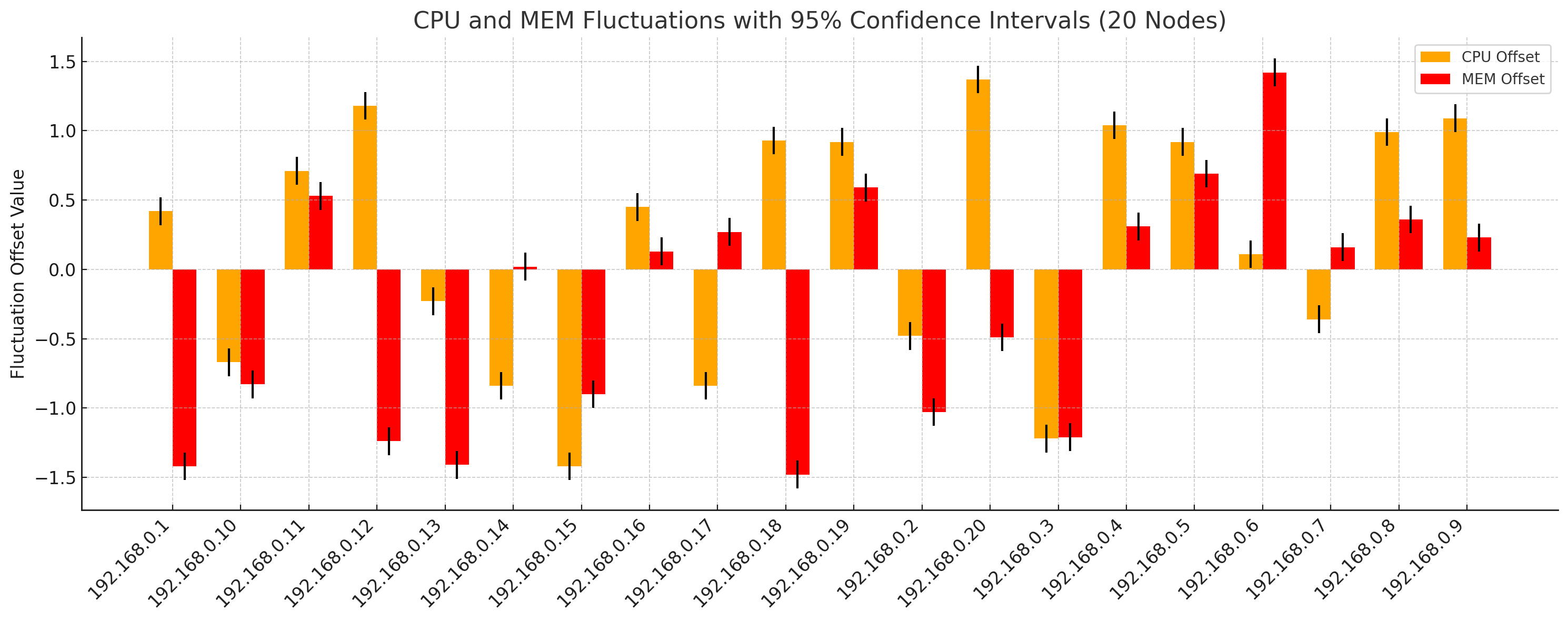}
    \caption{Example of fluctuation in node parameters of 20 nodes}
    \label{fig:fluctuations}
\end{figure}

This design avoids the problem of embedding collapse due to identical feature vectors being aggregated \cite{b27}. For every configuration step, GraphSAGE is executed and the resulting embeddings are extracted, normalized, and saved. The high-dimensional embeddings generated by our model contain rich structural and semantic information, but their complexity makes direct visualization impractical. Therefore, it is necessary to apply compression and dimensionality reduction techniques that can project these complex data into lower-dimensional spaces, while preserving their most significant features for analysis.

In this work, we employed PCA (Principal Component Analysis) 
as dimensionality reduction method. It was selected for its ability to identify directions of maximum variance, providing a linear and interpretable representation of the main components \cite{b28}. 

\subsection{Experiment: Semantic Drift of a Target Node as Local Knowledge Validation}
The experiment that we conducted, termed \textit{Semantic Drift of a Target Node}, aims to observe how a single node's embedding evolves in relation to a semantically stable environment as its own operational parameters are progressively altered.

We constructed three distinct network configurations, comprising \textit{5, 10, and 20 nodes}, each instantiated under three topological models: \textit{ring}, \textit{fully connected}, and \textit{chain}. Initially, the system is stabilized under a \textit{uniform workload distribution at 50\% of resource usage}, establishing a semantically rich \textit{baseline embedding space}. The embeddings generated in this phase encapsulate the equilibrium state in which all nodes have comparable system workload.

Subsequently, a \textit{target node} is selected, and its computational load is reset to 0\% and progressively increased from \textit{10\% to 100\%}, while the remaining nodes maintain their baseline workload. At each step, the target node refines its embedding through localized Knowledge Sharing, influenced by both its internal state and the semantic feedback from its neighbors.

The evolution of the target node's embedding is visualized through PCA (Fig. \ref{fig:line_results}, \ref{fig:ring_results}, \ref{fig:fully_results}), which projects the high-dimensional embeddings into a 2D latent space. Remarkably, the embedding trajectory for each of the analyzed network topologies delineates a \textit{curve} that diverges from the baseline in an interpretable manner. This geometric pattern is indicative of a semantic transformation that accurately reflects the incremental workload imposed on the node.

\begin{figure*}[ht!]
    \centering
    \begin{subfigure}[b]{0.22\linewidth}
        \centering
        \includegraphics[width=\linewidth]{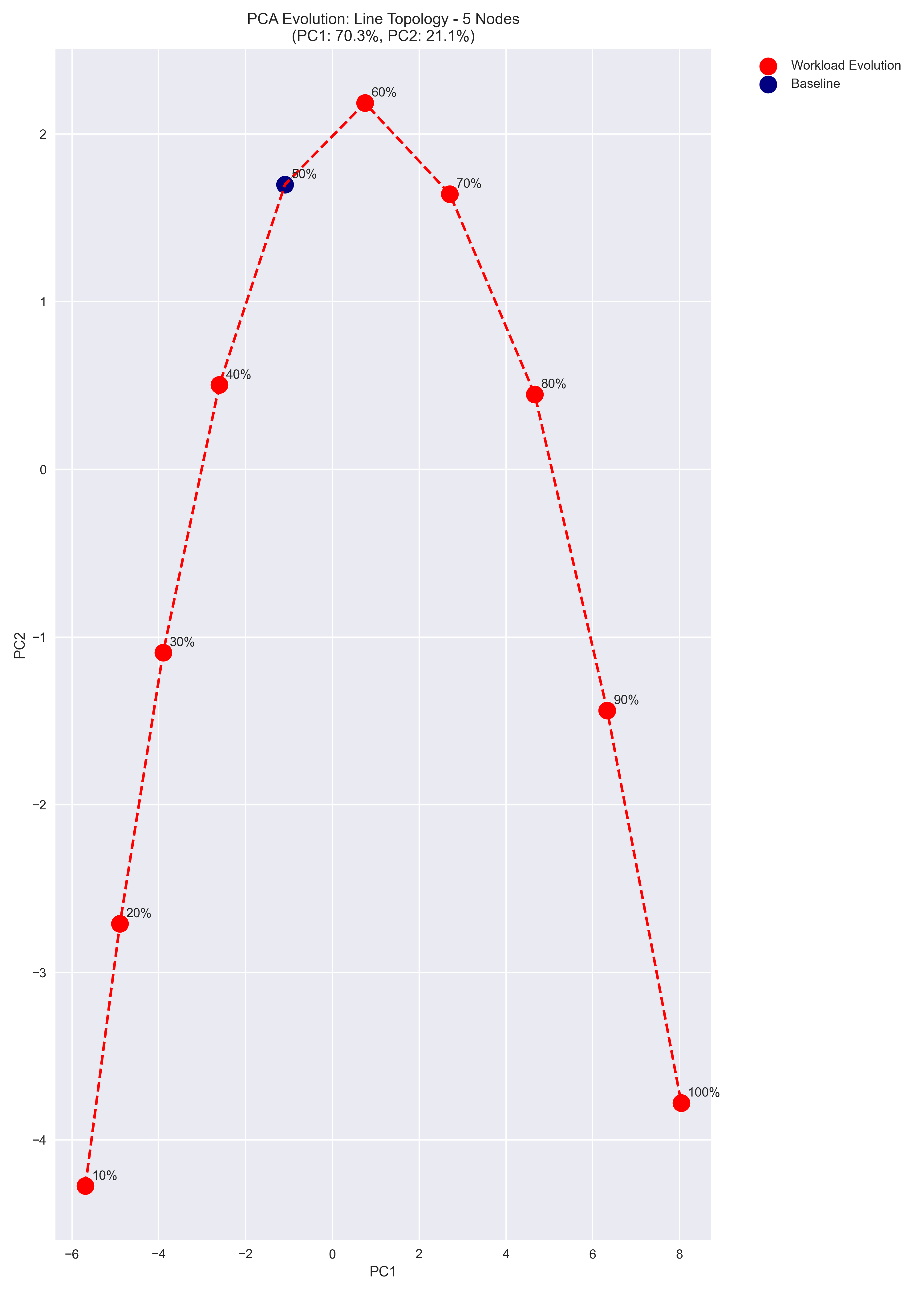}
        \caption{5 Nodes}
        \label{fig:line_5_nodes}
    \end{subfigure}
    \hfill
    \begin{subfigure}[b]{0.22\linewidth}
        \centering
        \includegraphics[width=\linewidth]{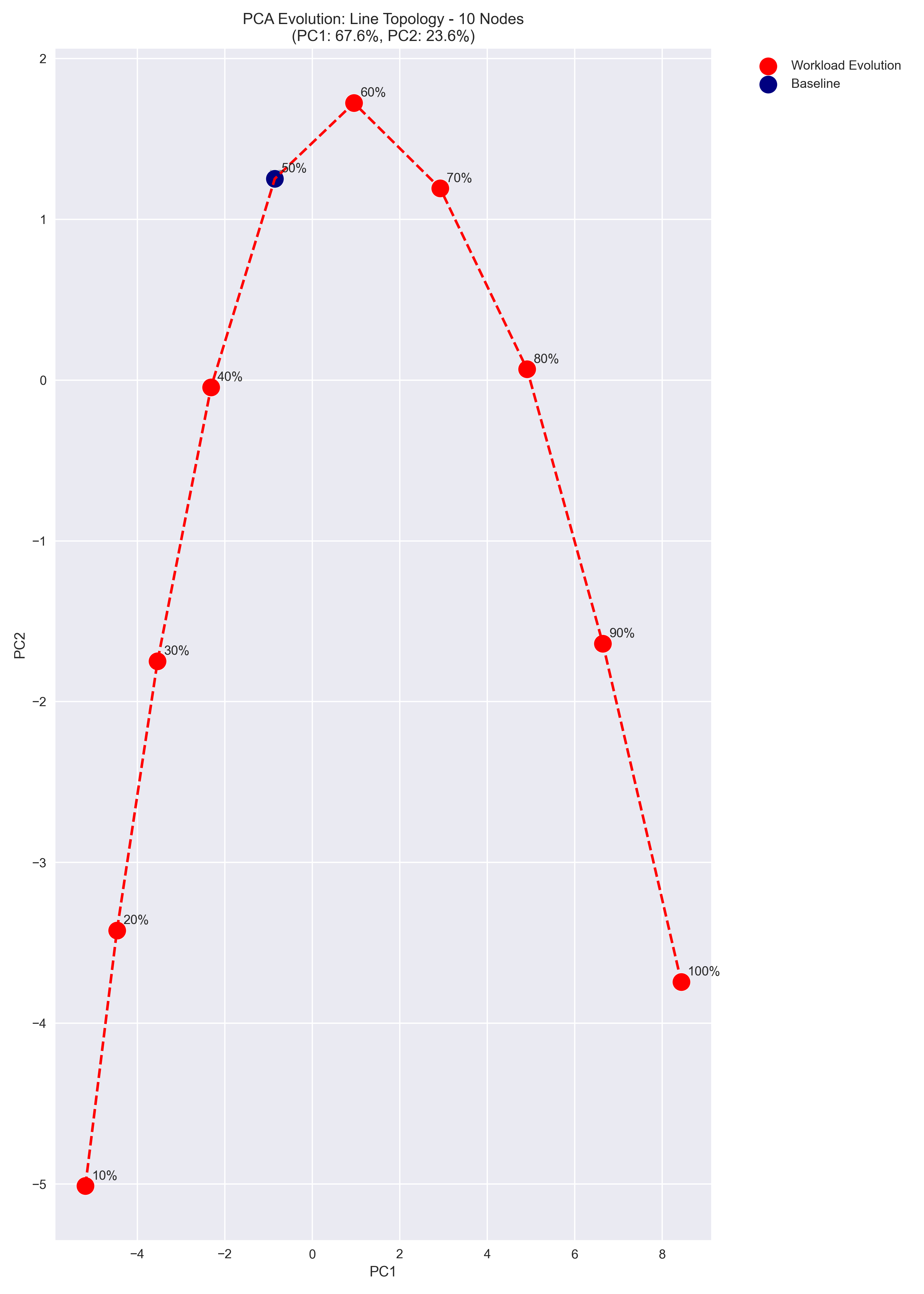}
        \caption{10 Nodes}
        \label{fig:fig:line_10_nodes}
    \end{subfigure}
    \hfill
    \begin{subfigure}[b]{0.22\linewidth}
        \centering
        \includegraphics[width=\linewidth]{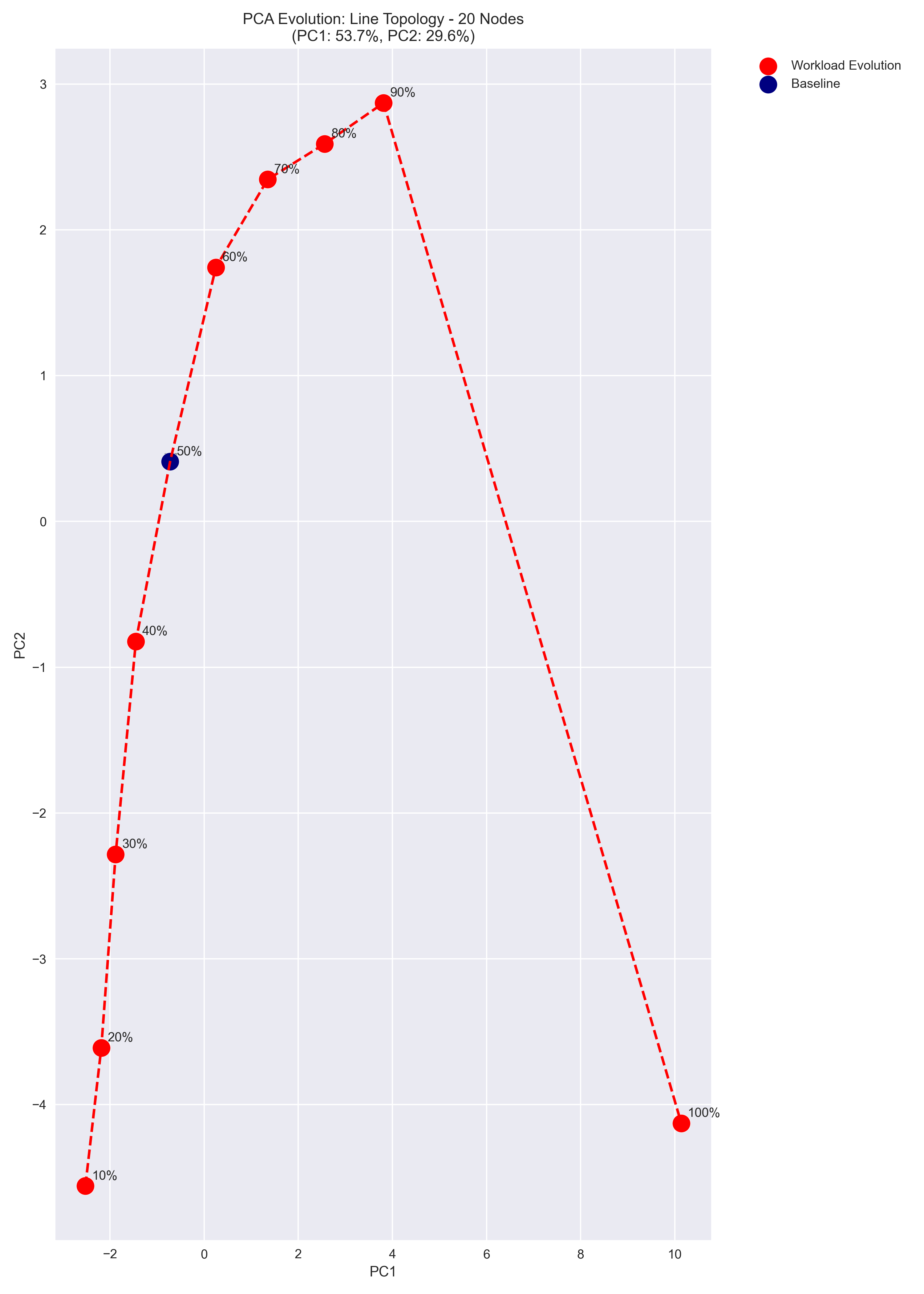}
        \caption{20 Nodes }
        \label{fig:line_20_nodes}
    \end{subfigure}
    \caption{PCA Workload visualization across different node configurations with Chain Topology.}
    \label{fig:line_results}
\end{figure*}

\begin{figure*}[ht!]
    \centering
    \begin{subfigure}[b]{0.22\linewidth}
        \centering
        \includegraphics[width=\linewidth]{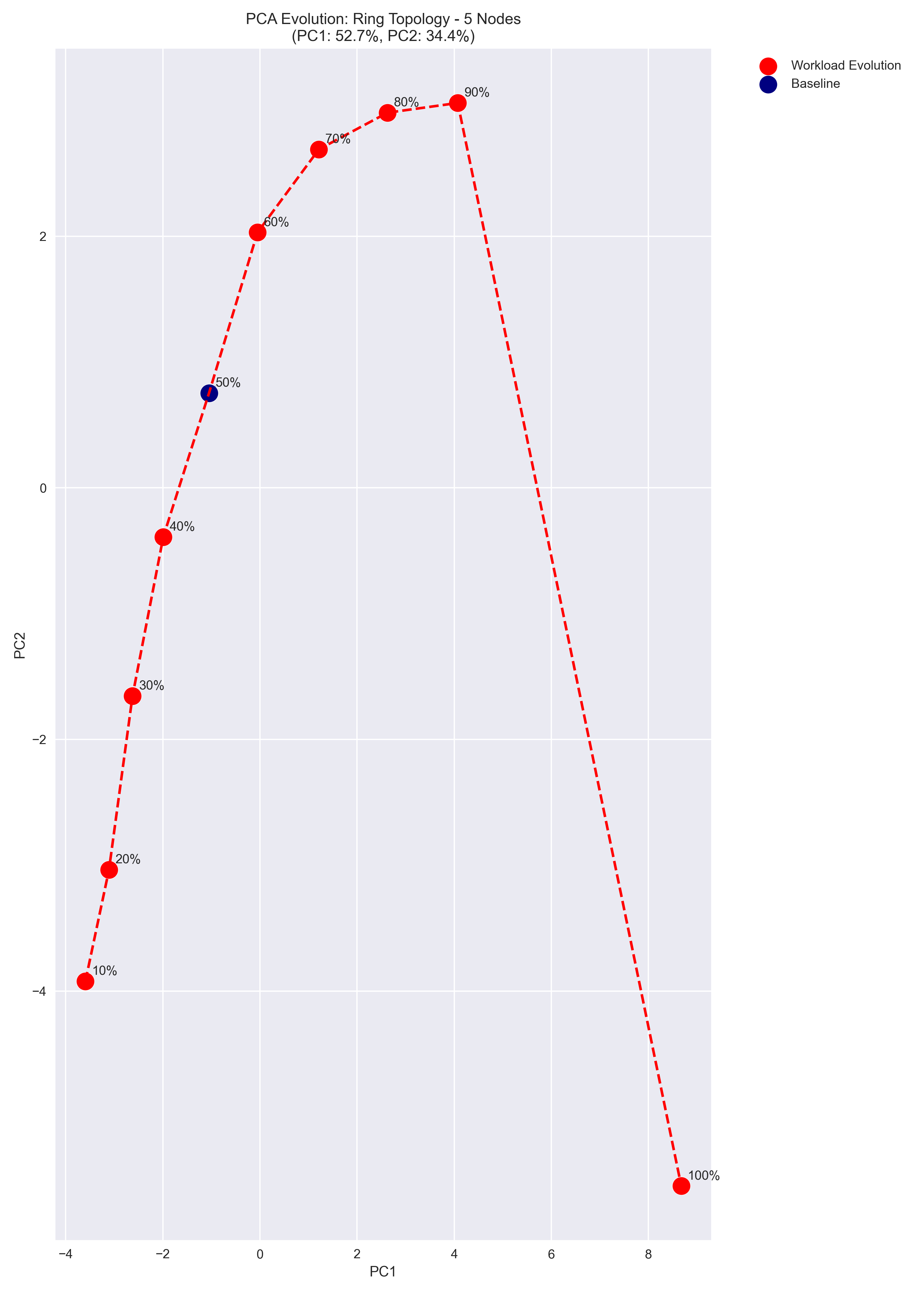}
        \caption{5 Nodes}
        \label{fig:ring_5_nodes}
    \end{subfigure}
    \hfill
    \begin{subfigure}[b]{0.22\linewidth}
        \centering
        \includegraphics[width=\linewidth]{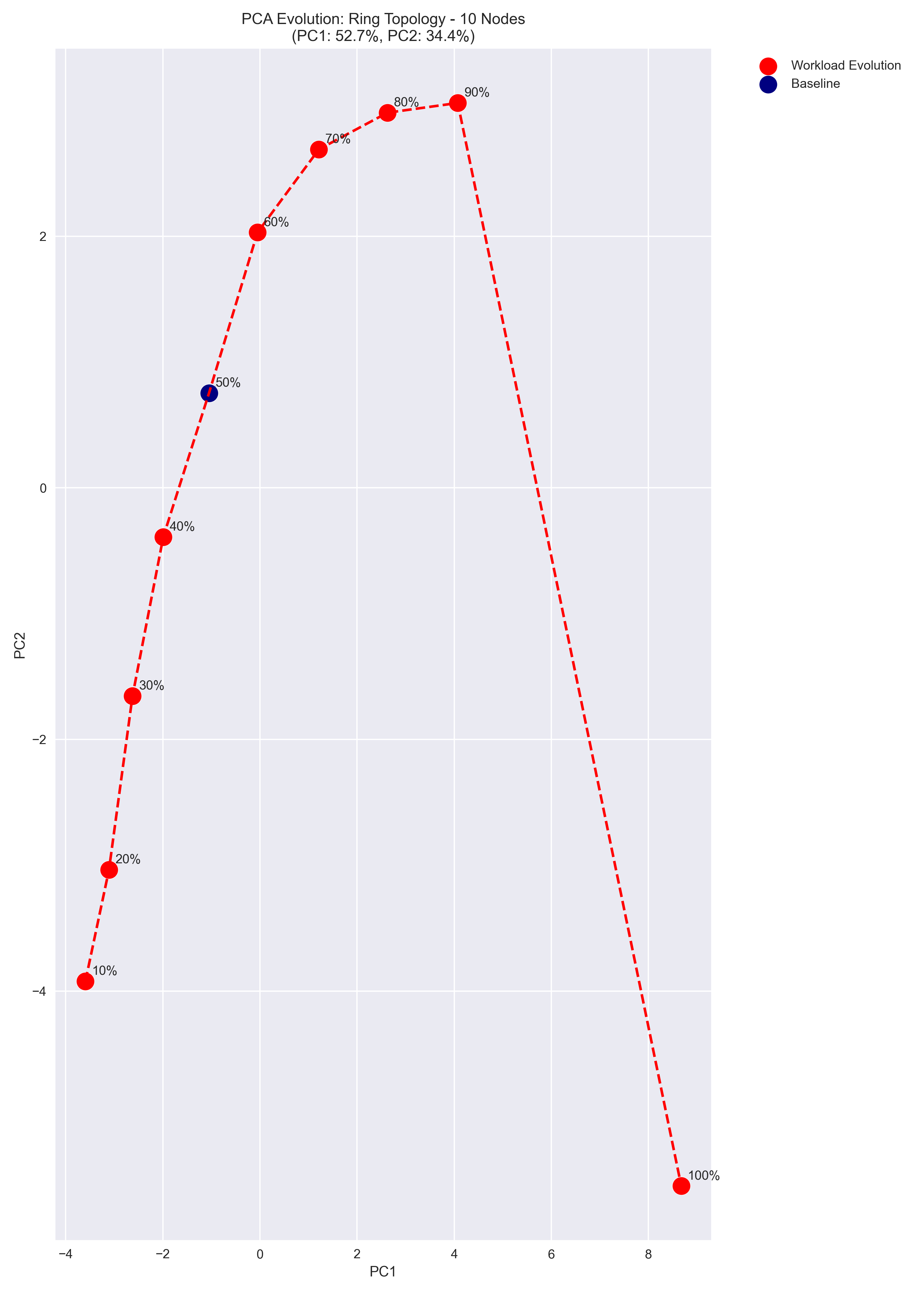}
        \caption{10 Nodes}
        \label{ffig:ring_10_nodes}
    \end{subfigure}
    \hfill
    \begin{subfigure}[b]{0.22\linewidth}
        \centering
        \includegraphics[width=\linewidth]{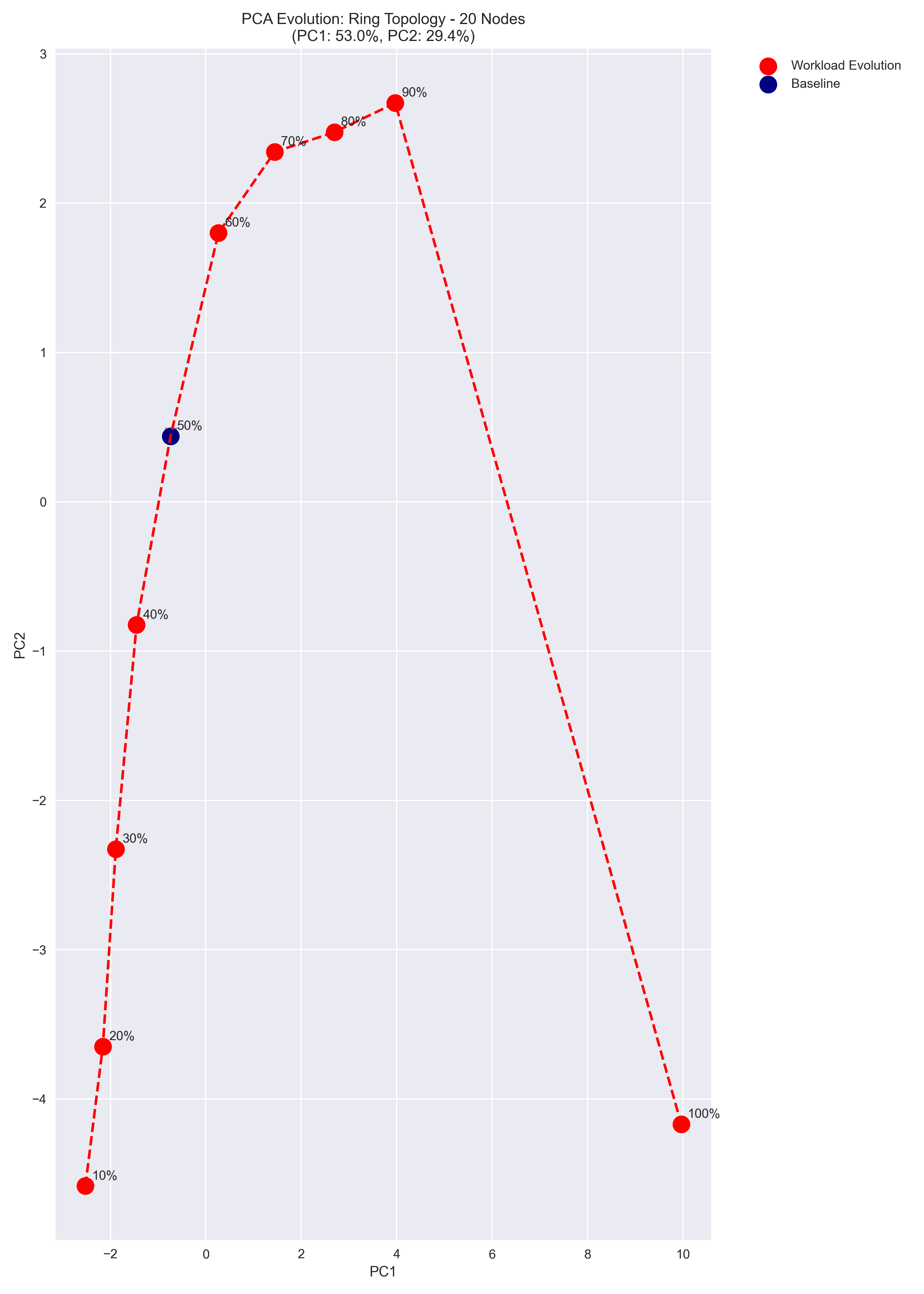}
        \caption{20 Nodes}
        \label{fig:fig:ring_20_nodes}
    \end{subfigure}
    
    \caption{PCA Workload visualization across different node configurations with Ring Topology.}
    \label{fig:ring_results}
\end{figure*}
\begin{figure*}[ht!]
    \centering
    \begin{subfigure}[b]{0.22\linewidth}
        \centering
        \includegraphics[width=\linewidth]{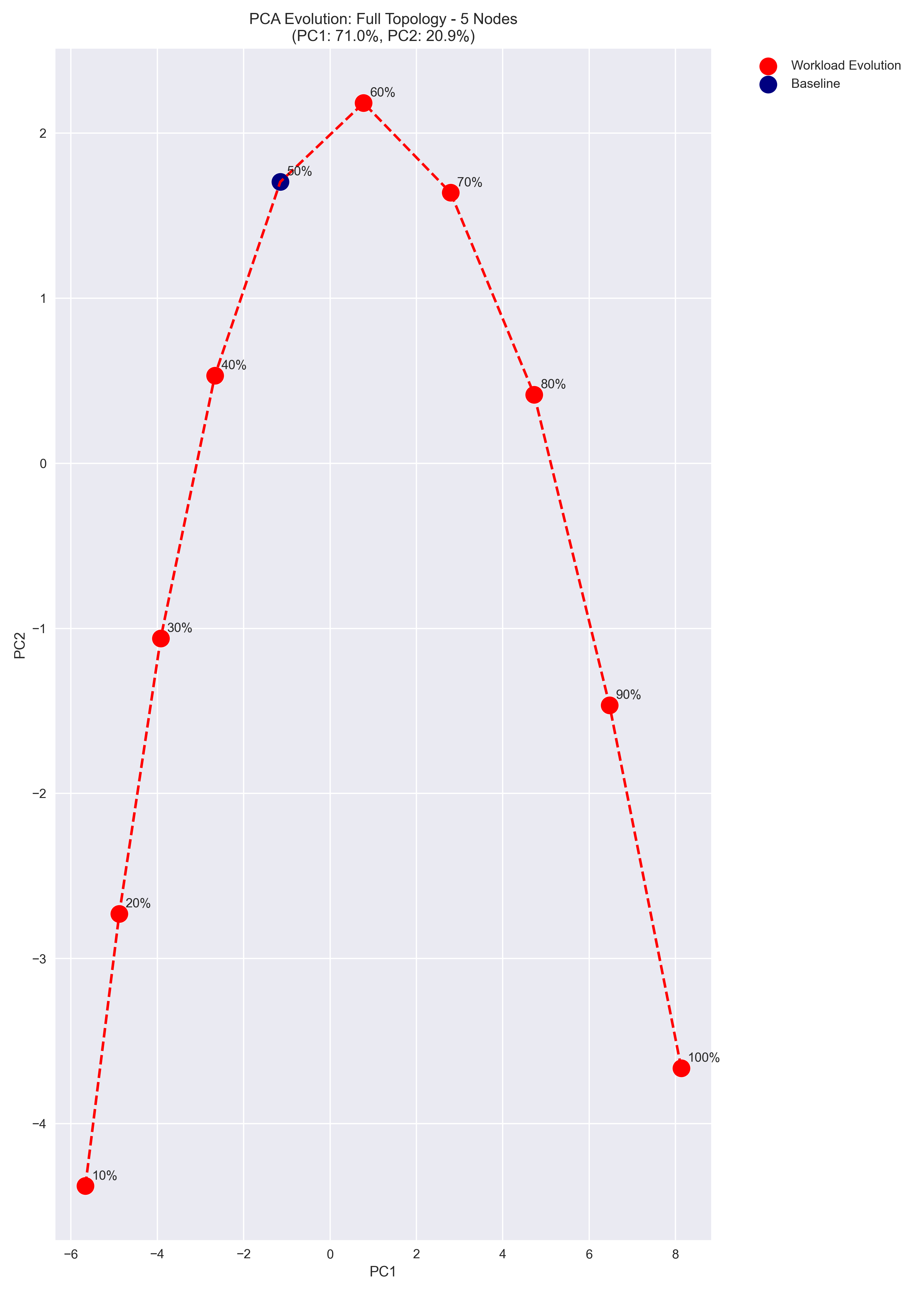}
        \caption{5 Nodes}
        \label{fig:fully_5_nodes}
    \end{subfigure}
    \hfill
    \begin{subfigure}[b]{0.22\linewidth}
        \centering
        \includegraphics[width=\linewidth]{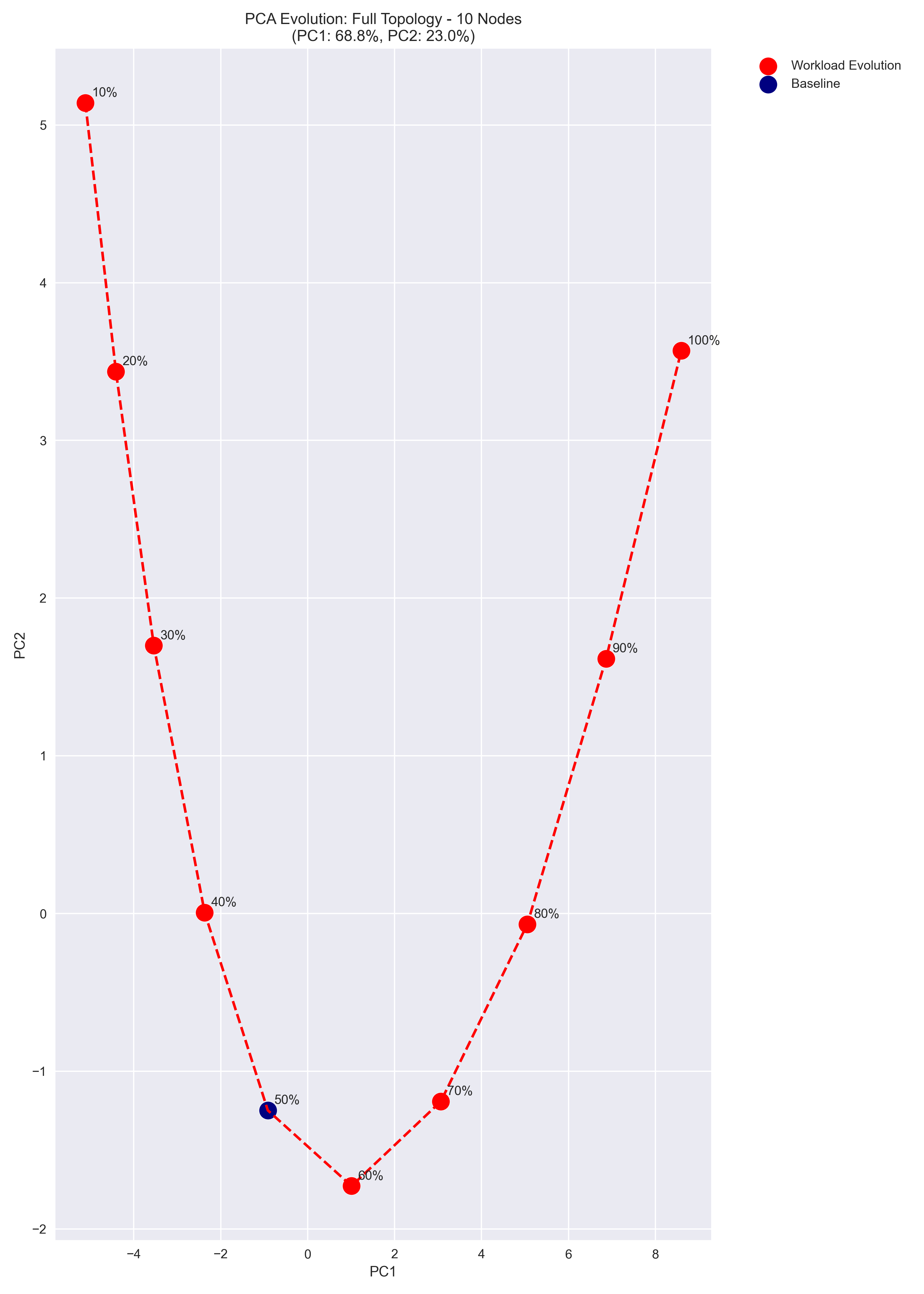}
        \caption{10 Nodes}
        \label{fig:fully_10_nodes}
    \end{subfigure}
    \hfill
    \begin{subfigure}[b]{0.22\linewidth}
        \centering
        \includegraphics[width=\linewidth]{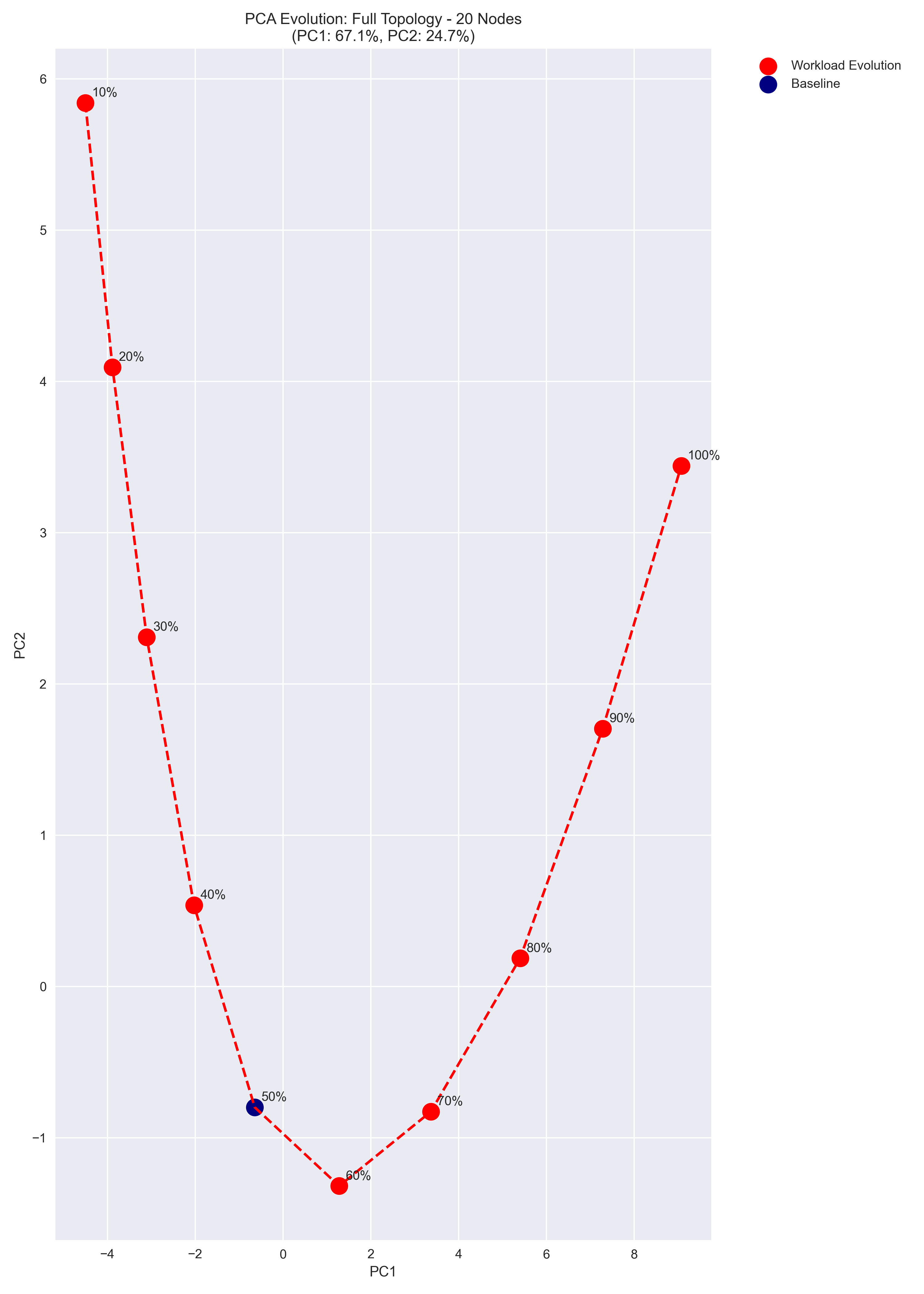}
        \caption{20 Nodes}
        \label{fig:fully_20_nodes}
    \end{subfigure}
    \caption{PCA Workload visualization across different node configurations with Fully Connected Topology.}
    \label{fig:fully_results}
\end{figure*}

Each embedding point along the trajectory serves as a semantic snapshot of the node's role and operational context, validating the premise that localized Knowledge Sharing fosters coherent and meaningful representations.

\section{Results Discussion}\label{sec:results_discussion}

The results obtained across the three topologies reveal how the structural properties of the network strongly influence the semantic drift of the target node. 

In the \textbf{chain topology} (Fig.~\ref{fig:line_results}), the trajectory of the target node follows a regular parabolic shape for all network sizes (5, 10, and 20 nodes). This behavior is explained by the sequential nature of knowledge propagation: the target node is influenced only by its immediate neighbors, and the embeddings progressively integrate information along the line. As the workload increases, the node diverges from the baseline but remains continuously attracted back by the stable cluster operating at 50\%. The resulting balance between local variation and global alignment generates the smooth, symmetric curve visible in the PCA space.

In the \textbf{ring topology} (Fig.~\ref{fig:ring_results}), the absence of boundaries modifies the dynamic significantly. Each node has two neighbors, but the information circulates symmetrically along the cycle. This leads to a more balanced diffusion of knowledge and reduces the dominance of any single direction of propagation. As a consequence, the trajectory of the target node appears smoother and less parabolic than in the line case, with a more stable central trend. The influence of the baseline cluster is still present, but it is uniformly distributed, resulting in a softened drift.

In the \textbf{fully connected topology} (Fig.~\ref{fig:fully_results}), the trajectory takes on a characteristic ``U'' shape. Here, each node directly exchanges information with all others, which amplifies the influence of the baseline. The target node is rapidly stabilized around the central state when its workload is close to 50\%, but diverges sharply at the extreme values of 0\% and 100\%. This symmetry reflects the immediate and global nature of knowledge propagation in a fully connected network, where any local deviation is instantly contrasted by the consensus of the entire cluster.

It is also worth noting that with a very small number of nodes (e.g., 5), the trajectory can appear inverted, resembling an upward parabola rather than the  ``U''. This effect is not due to a different semantic behavior but rather to the way PCA projects the embedding space: when the network is small, the target node exerts a proportionally stronger influence on the baseline cluster, and the main axis of variance may be oriented in the opposite direction. As the number of nodes increases (10 or 20), the global consensus becomes more stable, and the projection consistently reflects the symmetric ``U'' shape expected in a fully connected setting. In the line and ring topologies the knowledge propagation is either sequential or cyclic, which distributes the influence more gradually and uniformly, preventing the sharp symmetric ``U'' pattern observed in the fully connected case.

Overall, these findings demonstrate that the evolution of node embeddings is not determined only by the variation of individual workload, but is deeply conditioned by the underlying topology. Sequential propagation in the line emphasizes a parabolic drift, cyclic propagation in the ring distributes the influence and produces a smoother curve, while direct global connectivity in the fully connected case enforces a highly symmetric stabilization around the baseline.

\section{Conclusions And Future Works}\label{sec:conclusions}
In this paper, we introduced a novel Knowledge Graphs-Driven Intelligence framework for Distributed Decision Systems, addressing the inherent challenges of data heterogeneity, dynamic environments, and the critical need for decentralized coordination in modern distributed systems. Our innovative approach makes use of the semantic richness of Knowledge Graphs (KGs) and the representational power of Graph Embeddings (GEs) to achieve decentralized intelligence through Knowledge Sharing.

We presented a comprehensive 4-layer architecture comprising the Physical Layer, Storage Layer, Knowledge Layer, and Decision Layer. This architecture empowers individual nodes to locally construct and continuously refine semantic representations of their operational context. The core of our system lies in the Knowledge Sharing mechanism, where nodes iteratively aggregate embeddings through neighbor-based exchanges using GraphSAGE. This iterative local aggregation process dynamically evolves into a global semantic abstraction, which we named Knowledge Map, enabling coordinated decision-making without centralized control.

To validate our approach, we first provided preliminary experimental evaluations within the context of a distributed orchestration use case, simulating various KG topologies and varying operational conditions. The Semantic Drift experiment empirically showed how individual nodes refine their embeddings in response to local operational changes, maintaining semantic alignment. These results consistently confirmed that the distributed knowledge-sharing mechanism effectively maintains semantic coherence, adaptability, and scalability, making it highly suitable for complex and dynamic environments such as Edge Computing, IoT, and multi-agent systems.

Future work will include a rigorous mathematical validation of the proposed model, extending beyond the preliminary evaluations presented here. In addition, we plan to perform the Embedding Separation experiment to quantitatively and qualitatively assess the system's ability to distinguish between heterogeneous configurations, thus validating the emergent coherence of the shared Knowledge Map. We also intend to extend the evaluation by considering both homogeneous and heterogeneous datasets, in order to demonstrate the generalizability of the approach across diverse operational scenarios. These extensions will be addressed in the full version of the work, providing a more complete and robust validation of the proposed framework.

\begin{acks}
Rosario Napoli is a PhD student enrolled in the National PhD in Artificial Intelligence, XL cycle, course on Health and
life sciences. This work has been partially funded by 
the Horizon Europe ``Open source deep learning platform dedicated to Embedded hardware and Europe'' project (Grant Agreement, project 101112268 - NEUROKIT2E),
the “SEcurity and RIghts in the CyberSpace (SERICS)” partnership (PE00000014), under the MUR National Recovery and Resilience Plan funded by the European Union – NextGenerationEU. In particular, it has been supported within the SERICS partnership through the projects FF4ALL (CUP D43C22003050001) and SOP (CUP H73C22000890001), the Italian Ministry of University and Research (MUR) ``Research projects of National Interest (PRIN-PNRR)'' through the project “Cloud Continuum aimed at On-Demand Services in Smart Sustainable Environments” (CUP: J53D23015080001- IC: P2022YNBHP) and the Italian Ministry of University and Research (MUR) ``Research projects of National Interest (PRIN)'' call through the project ``Tele-Rehabilitation as a Service (TRaaS)'' (CUP J53D23007060006). 
\end{acks}

\bibliographystyle{ACM-Reference-Format}

\bibliography{bibliography}


\end{document}